\documentclass[twocolumn,showpacs,preprintnumbers,amsmath,amssymb]{revtex4}

\usepackage{graphicx}
\usepackage{dcolumn}
\usepackage{bm}

\begin{document}

\title{Charge-induced maximal spin states of a polynuclear  transition-metal complex}

\author{C. Romeike$^1$, M. R. Wegewijs$^1$, M. Ruben$^2$, W. Wenzel$^2$, and H. Schoeller$^1$} 
\affiliation{
$^{1}$ Institut f\"ur Theoretische Physik A, RWTH Aachen, 52056 Aachen,  Germany \\
$^{2}$Institut f\"ur Nanotechnologie, Forschungszentrum Karlsruhe,  76021 Karlsruhe, Germany}

\date{\today}

\begin{abstract}
  We theoretically investigate the ground state spin of a polynuclear
  transition-metal complex as a function of the number of added
  electrons  taking into account strong electron correlations.
  Our phenomenological model of the so-called [$2\times2$]-grid
  molecule incorporates the relevant electronic degrees of freedom on the four
  transition-metal centers (either Fe$^{2+}$ or Co$^{2+}$)
  and the four organic bridging ligands. Extra electrons preferably occupy  redox orbitals on the ligands.
  Magnetic interactions between these ligands are mediated by
  transition-metal ions {\em and vice versa}. Using both perturbation theory and exact diagonalization
  we find that for certain charge states the maximally attainable total
  spin (either $S_{\mathsf{tot}}=3/2$ or $S_{\mathsf{tot}}=7/2$) may actually be achieved. Due to the Nagaoka mechanism,
  all unpaired electron spins couple to a total maximal spin, including unpaired
  electron spins  on the metal-ions in the case of Co$^{2+}$.
  The parameters are chosen to be consistent with cyclovoltammetry
  experiments in which up to twelve redox states have been observed.
  The above effect may also be realized in other complexes with an
  appropriate connectivity between the redox sites. The maximal spin states of such a charge-switchable molecular magnet
  may be experimentally observed  as spin-blockade effects on the
  electron  tunneling in a three-terminal transport setup.
\end{abstract}
\maketitle
\section{Introduction}
\begin{figure}
 \begin{center}
  \includegraphics[angle=0,scale=0.2]{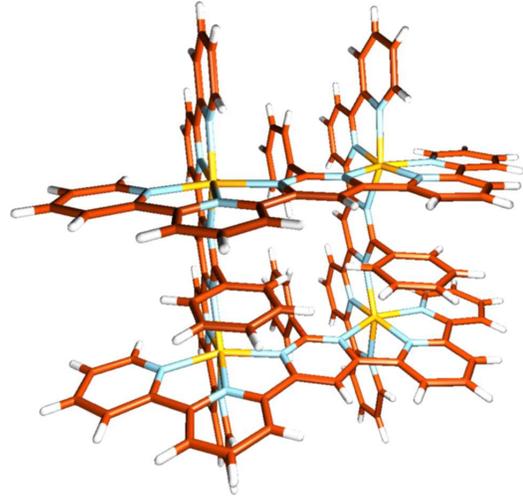}
  \caption{Structure of the [$2 \times 2$]-grid-type complex.}
  \label{fig:complex}
 \end{center}
\end{figure}

The synthesis and investigation of molecule based magnets have been
active areas of research for almost a decade~\cite{sessoli93,aubin99}. 
Today molecule based magnets offer a 
variety of chemical and magnetic properties which have potential
applications in a wide range of systems.  
Photo-magnetic (e.g. K$_{0.4}$Co$_{1.3}$[Fe(CN)$_6$]) or spin-crossover magnetic substances
(e.g.  Fe(o-phenanthroline)$_2$(NCS)$_2$) offer potential applications
in switchable devices. 
Polynuclear transition-metal complexes are particularly attractive
in the effort to design molecules with magnetic centers which couple
 either ferro- or antiferromagnetically.
Interesting quantum tunneling effects were reported for
ferromagnetic  molecules,
e.g. Mn$_{12}$~\cite{friedman96} and
Fe$_8$~\cite{wernsdorfer99}.  
Antiferromagnets are of interest also due to effects associated with
the N{\'e}el-vector~\cite{chiolero98,meier01,waldmann03}.
One interesting class of highly-designable
complexes is the [$M \times M $]-grid structure which is formed by
self-assembly~\cite{lehn95,hanan97,ruben04,zhao00}.
It consists of $M^{2}$ transition-metal centers which are positioned by two perpendicular
arrays of rod-like ligands, each with $M$ coordination sites.
The transition-metals typically lie in a plane due to symmetric
positioning of the ligands.
Several realizations of these supramolecules exist which exhibit a
variety of magnetic and electrochemical properties, which can be
manipulated via sidegroup substitutions. In this paper we focus in
particular on the [$2\times2$]-grid depicted in Figure
\ref{fig:complex}. 
It consists of four organic ligands
(bis(bipyridyl)-bipyrimidine) and four transition-metal ions
(e.g. Mn$^{2+}$, Fe$^{2+}$, Co$^{2+}$, Ni$^{2+}$ or Zn$^{2+}$
\cite{ruben03a}). Each metal-ion is situated in an approximately
octahedral environment of nitrogen atoms. The total charge 8+ of the
complex is countered by anions, e.g. $\mbox{BF}_{4}^{-}$.  The
Co$^{2+}$-[$2\times2$]-grid exhibits an impressive twelve reversible
reduction steps in solution in a cyclic voltammetry at $T=253 K$
\cite{ruben03a}.  It is also possible to trap and identify these
complexes on a graphite surface in a controlled way~\cite{semenow99b}.
Furthermore, in Co$^{2+}$-[$2\times2$]-grids the spin of each
metal-ion was found to be $S=1/2$ at low temperature and their
intramolecular magnetic coupling was demonstrated to be
antiferromagnetic~\cite{waldmann97}.\\ 
We have developed a phenomenological strongly-correlated electron model which
represents the important spin and charge degrees of freedom that play
a role in both the redox chemistry and the magnetic interactions of
[$2\times2$]-grid complexes.
 We include the strong electrostatic
interactions between electrons localized on the metal-ions and the
ligands as well as the weak tunneling of electrons between them.
  A crucial feature of our model is that the orbital and charging energies on the
ligands and ions are very different, a situation common
in transition-metal complexes.
We are interested in how the electron spins on the molecule couple to form
the total magnetic moment and how this coupling is affected by total
electron number which can be electrically controlled in a three
terminal transport setup (which is not considered here).
The metal-ions mediate the coupling between electrons on the ligands
{\em and vice versa}.
We also want to study the effect of an unpaired electron spin on a
mediating metal-ion on the ligand-ligand spin coupling, which will in
turn determine the total spin formed by the ion and ligand subsystems.
Two simple cases are therefore considered, assuming each ion is in a low-spin state:
either $S=0$, Fe$^{2+}$, or $S=1/2$, Co$^{2+}$.
 The Coulomb interaction also generates direct exchange coupling
 between unpaired electrons on ligands and metal-ions which we also
 take into account. We focus on how the spins couple to a magnetic moment of large magnitude
 due to strong Coulomb and kinetic effects depending on the total charge.
 We do not consider effects on smaller energy scales 
 which determine the preferred direction of the magnetic moment.
\\
As expected, an antiferromagnetic coupling between the
 metal-ion and ligand subsystems prevails for most total charge numbers.
Extra electrons added to the bridging ligands introduce a
ferromagnetic spin correlation between the metal-ions if the latter have a spin. 
However, we find that near half-filling of the four
ligand orbitals sufficiently strong Coulomb
interactions can stabilize a {\em maximal spin ground state}
($S_{\mathsf{tot}}=3/2$ for Fe$^{2+}$ and $7/2$ for Co$^{2+}$) by the
Nagaoka mechanism~\cite{nagaoka66}. Due to the Pauli principle a
missing or excess electron on the ligands (relative to half-filling)
can be delocalized maximally when the ``background'' of the remaining
electrons is completely spin-polarized. In our model this effect is
most easily achieved for a Fe$^{2+}$-[$2\times2$]- grid. For a
Co$^{2+}$-[$2\times2$]-grid the spins of electrons localized on the
bridging metal-ions counteract the Nagaoka mechanism.  However, the
direct exchange coupling between unpaired electrons on the neighboring
ligands and metal-ions cooperates with the Nagaoka
mechanism~\cite{kollar96}.
The resulting maximal spin is however more than twice as large as for
Fe$^{2+}$. The charge-sensitive spin-polarization requires a strong local
Coulomb charging effect and metal-to-ligand
charge-transfer (MLCT) barrier.
The former can be achieved chemically by introducing electron-donating
sidegroups on the ligands~\cite{ruben03a}. The electrochemical experiments~\cite{ruben03a} were performed near
room temperature where thermal occupation of high-spin excited states
of the metal-ions becomes possible. Generalization of our model to
incorporate the high-spin states of the ions is, however, nontrivial
since vibrational degrees of freedom of the nuclear framework are
involved in their stabilization~\cite{guetlich94}.
This is beyond the scope of the present paper
 where we focus on low temperature behavior.
\\
Experimental detection of sublattice and total magnetization as a
function of the number of added electrons would be of great interest.
In particular, we propose electron transport through the complex
in a three terminal setup.
In a gate voltage range where the ligand orbitals are near
half-filling the large change in ground state spin
\big($\Delta S_{\mathsf{tot}}=3/2$ (Fe$^{2+}$), resp. $\Delta
S_{\mathsf{tot}}=7/2$ (Co$^{2+}$) \big) may be revealed through the
\emph{spin-blockade} effect~\cite{weinmann95}. 
We have investigated this transport problem in
detail for a simple generic model n~\cite{romeike05a}.
The goal of the present paper is to derive and investigate the
effective model, in particular the effect of bridging metal-ions with a spin. Experiments in semiconductor heterostructures
 show that a transport current can couple through
 the spin-blockade effect to electron spins localized in a quantum dot
and subsequently to nuclear spins in the local environment via the
hyperfine interaction~\cite{ono03}.
This offers an interesting perspective for single-molecule
transport where the immediate nuclear environment of the device can be
controlled at the level of chemical synthesis.
\\
 The paper is organized as
follows. In section~\ref{sec:El-struct} we discuss the minimal set of
orbitals involved in electron addition processes and the
intramolecular magnetic coupling. This serves as a motivation for the
correlated electron model which we introduce in section
\ref{sec:model}. In section~\ref{sec:pt} we discuss a perturbative
treatment of the ion-ligand tunneling in this model to gain a simple
understanding of the role of spin degrees of freedom on the ions. We
obtain the ground states and low energy total-spin excitations in sections
\ref{sec:specFe} and~\ref{sec:specCo}. We discuss the electron
addition energies and the modulation of the ground state spin as a
function of the number of added electrons.  We conclude with a
discussion of implications for electron tunneling experiments in
section~\ref{sec:disc}.
\section{Electronic structure\label{sec:El-struct}}
\begin{figure}
 \begin{center}
  \includegraphics[angle=-90,scale=0.3]{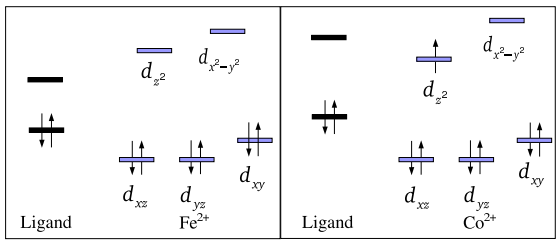}
  \caption{Orbital configuration of transition metal-ions and ligands.}
  \label{fig:orb_config}
 \end{center}
\end{figure}

\begin{figure}
 \begin{center}
  \includegraphics[angle=270,scale=0.4]{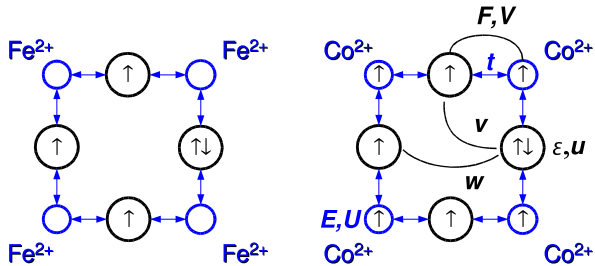}
  \caption{Geometry of transition-metal [$2\times2$]-grid consisting
    of metal-ion orbitals (small circles), Fe$^{2+}$ (left) and
    Co$^{2+}$ (right), connected by ligands (large circles). The hopping
    $t$, orbital energies $E, \epsilon$ and Coulomb charging $U, V, u,
    v, w$ and direct exchange energy $F$ are shown schematically.}
  \label{fig:geometry}
 \end{center}
\end{figure}

We can expect that the dominant molecular orbitals (MOs) at the Fermi
energy are either $d$-like orbitals from the metal 2+ ions or
$\pi$-orbitals from the ligands. The electrochemical experiments
\cite{ruben03a} indicated that additional electrons occupy orbitals
localized mostly on the \emph{ligands}, which is not uncommon for
polypyridine complexes~\cite{vlcek82}.\\ Let us first consider the individual 
metal-ions which are coordinated by six nitrogen atoms. In a ligand-field 
picture, an octahedral coordination of each
metal-ion in the grid would cause their $d$-orbitals to split up into
two shells, $t_{2g}$ ($d_{xy}, d_{xz},d_{yz}$) and $e_{g}$
($d_{x^{2}-y^{2}}, d_{z^{2}}$). However, the symmetry of the local
environment is lower and results in a small splitting of the levels in
each shell. The different orbital occupations for Fe$^{2+}$ and
Co$^{2+}$ are shown in Figure~\ref{fig:orb_config}. 
In the case of Fe$^{2+}$ in its low-spin configuration the $t_{2g}$
 shell is closed and the $e_{g}$ orbitals are empty.  In the case of
 Co$^{2+}$ there is an additional unpaired electron in the lowest of
 the two $e_{g}$ orbitals.\\ The highest occupied molecular orbitals
 of the ligands are mostly of $\pi$ character as ab-initio
 calculations show. The important point is that the empty or singly
 occupied $e_{g}$-orbitals on the metal-ions have different
 approximate symmetry ($\sigma$).  This means we can consider ligands
 and metal-ions weakly coupled systems and discuss the tunneling of electrons
 between them.  The remaining question is whether extra
 electrons prefer to occupy the metal-ion $d$-orbitals or the ligand
 $\pi$-orbitals. As far as charging effects are concerned, one expects that the onsite
energy $U$ on the metal ions is larger than the on-site energy $u$ on
the ligands because orbitals are more contracted on the metal-ions.
Furthermore, electrons would favor positions between two positively
charged ions, i.e. on the bridging ligand. This is in line with the
interpretation of the cyclovoltammetry experiments~\cite{ruben03a}.
The main difference in our model between the Co$^{2+}$/Fe$^{2+}$-[$2\times2$]-grids 
will thus be the presence/absence of an
localized, unpaired electron in an orbital of the metal-ion bridging
the two ligands (Figures~\ref{fig:orb_config},~\ref{fig:geometry}).
\section{Charging model\label{sec:model}}
We model the [$2\times2$]-grid complex by four metal (Fe$^{2+}$ or
Co$^{2+}$) and four ligand sites with one spin-degenerate orbital per
site (Figure~\ref{fig:geometry}). The following Hamiltonian then
captures the features of the electronic degrees of freedom discussed
in section~\ref{sec:El-struct}:
\begin{eqnarray}
  \label{eq:ham}
     H &=&  H_{\mathsf{T}} + H_{ \mathsf{L}} + H_{ \mathsf{Fe/Co}}+ H_{\mathsf{dir}}  +  H_{\mathsf{V}},\\
   \label{eq:ham_tun}
     H_{\mathsf{T}} &=&  \sum_{\langle i,j\rangle}\sum_{\sigma} t\, A^{\dag}_{i, \sigma} a_{j,
               \sigma} + h.c.\\
  \label{eq:ham_lig}
     H_{ \mathsf{L}} &=& \sum_{j=1}^{4}( \epsilon   n_{j} + 
                  u\, n_{j, \uparrow}  n_{j,\downarrow}  + 
                  v\, n_{j}   n_{j+1} ) \nonumber\\ &+&  w \sum_{j=1}^{2}  n_{j}   n_{j+2}  \\
  \label{eq:ham_met}
     H_{\mathsf{Fe/Co}} &=& \sum_{i=1}^{4}( E  N_{i} +   U\, N_{i, \uparrow}  N_{i, \downarrow})\\
  \label{eq:dir_x}
    H_{\mathsf{dir}} &=& F  \sum_{\sigma, \sigma'} \sum_{\langle i,j\rangle}
                         A_{i,\sigma}^{\dag} a_{j,\sigma'}^{\dag}
                         A_{i,\sigma'} a_{j,\sigma}  \nonumber\\ 
                    &=& -2 F \sum_{\langle i,j\rangle} (\mathbf{S}_{i} \mathbf{s}_{j,j}
                          +\frac{1}{4}\, N_{i}\,n_{j})\\
   \label{eq:V}
      H_{\mathsf{V}} &=& V \sum_{\langle i,j\rangle} N_{i}\,n_{j}.
\end{eqnarray}
Operators and variables (except $t, F, V$) in lower/upper case relate to the
metal-ion/ligands and it is implicitly understood that all indices
appearing run from 1 to 4 (e.g. $j+1\rightarrow 1$ for $j=4$). $\langle i,j\rangle$ denotes a summation over nearest
neighbor metal-ions ($i=1-4$) and ligands ($j=1-4$). 
The Fermion operator $a^{\dag}_{j, \sigma}\, (a_{j, \sigma})$ creates
(destroys) an electron on ligand site $j = 1-4$ with spin projection $\sigma = \pm
1/2$. The occupation number operator is defined as usual  $n_{j,\sigma} =  a^{\dag}_{j,
  \sigma} a_{j, \sigma}$ and $n_{j} = \sum_{\sigma}
n_{j,\sigma}$. Similar definitions hold for the metal-ion ($ A_{i,
\sigma}, N_{i, \sigma} = A^{\dag}_{i, \sigma} A_{i,
\sigma},N_{i}=\sum_{\sigma} N_{i,\sigma}$).
 $\boldsymbol{S}_{i} = \frac{1}{2} \sum_{\sigma, \sigma'} A^{\dag}_{i,
     \sigma} \boldsymbol{\tau}_{\sigma, \sigma'}  A_{i,\sigma'}$,
   where $\boldsymbol{\tau}$ is the vector of Pauli matrices, is the electron spin of a metal-ion and  
 $\mathbf{s}_{j,k} =   \frac{1}{2} \sum_{\sigma, \sigma'}a^{\dag}_{j,
   \sigma} \boldsymbol{\tau}_{\sigma, \sigma'}  a_{k, \sigma'}$, is an operator related to the ligands.
 The tunneling term (\ref{eq:ham_tun}) describes hopping between
ligand and metal-ions. For a $D_{2d}$ symmetric molecular structure
the hopping matrix elements $t$ are independent of sites. In the following we choose $t=1$ and all energies are 
expressed in units of $t$. The ligand-part of the Hamiltonian in (\ref{eq:ham_lig}) consists of a
spin independent orbital energy $\epsilon$, the classical Coulomb
repulsion terms on the ligand ($u$) and between adjacent ($v$) and the
opposite ligands ($w$).  Due to decreasing overlap with distance we
have $u > v/2 \geq w$ (e.g. for Fe$^{2+}$ $u \approx 4v \approx 0.3 eV$
and $v \approx 2w$~\cite{ruben03a}).  Equation (\ref{eq:ham_met})
describes the isolated metal-ion orbitals with energy $E$.  For the
metal-ions we only consider the short-range interaction $U$ because
the orbital overlap between two ions is much smaller than that between
two ligands. In order to describe the spin coupling of metal-ion and
ligand electrons correctly, we include the direct exchange
(\ref{eq:dir_x}) with $F>0$, which is known to stabilize ferromagnetic
states even when it is \emph{weak}~\cite{kollar96} (see sections
\ref{sec:pt} and~\ref{sec:Results}).  For consistency the metal-ligand
charging energy (\ref{eq:V}) also needs to be incorporated. In general
$F \lesssim V$ are of the same order. However, in the regime of
interest the particle number on the metal-ions is fixed ($N=0$ for
Fe$^{2+}$, $N=4$ for Co$^{2+}$, cf. section~\ref{sec:El-struct}), so
the second term in (\ref{eq:dir_x},\ref{eq:V}) yield a constant. In
Figure~\ref{fig:geometry} these interactions are schematically
indicated.\\
We study the parameter regime where the first eight extra electrons
 will occupy the ligands as in the experiment~\cite{ruben03a}. (To
 describe more than 8 reduction steps more than one orbital per ligand
 has to be taken into account. This is not our purpose here.) For
 Fe$^{2+}$ we must then assume that the two charge states of the
 ligand lie below the Fe$^{2+}$ orbital energy $E$:
\begin{eqnarray}
\mbox{Fe$^{2+}$}: \epsilon < \epsilon +u < E < E+U
\end{eqnarray}
For Co$^{2+}$ the two charge states of the ligand lie \emph{between}
the singly and doubly occupied states of the metal-ion:
\begin{eqnarray}
\mbox{Co$^{2+}$}: E < \epsilon < \epsilon +u < E+U
\end{eqnarray}
The orbital energy difference $\Delta = \epsilon -E$ is associated
with metal-ion to ligand charge transfer (MLCT) between unoccupied
metal-ion and ligand sites. In the case of Fe$^{2+}$ $\Delta < 0$ and
in the case of Co$^{2+}$ $\Delta > 0$.
\section{Perturbation theory - Effective model\label{sec:pt}}
\begin{figure}
 \begin{center}
  \includegraphics[angle=270,scale=0.3]{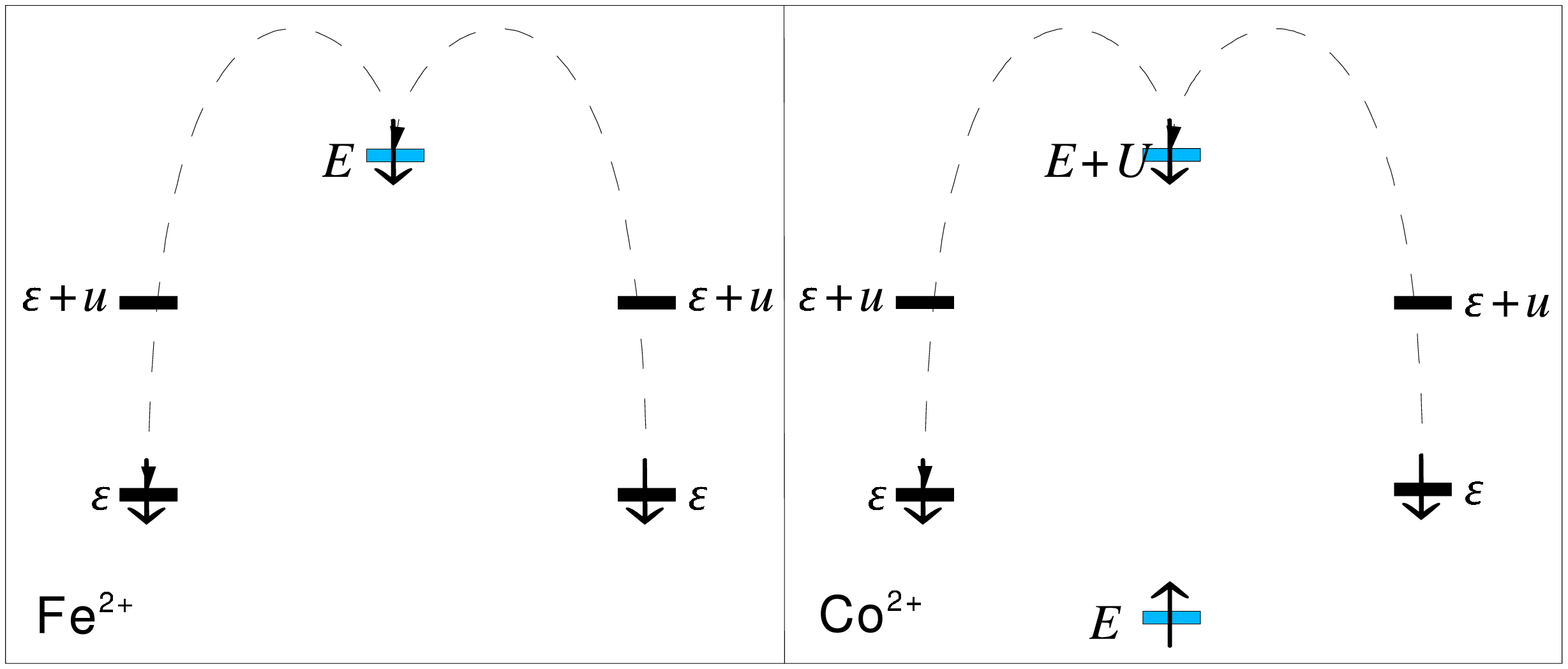}
  \caption{Energy diagram: Example of a charge transfer from one
    ligand to another ligand by virtual occupation of
    Fe$^{2+}$/Co$^{2+}$-ion state in the middle. Note that the MLCT
    barrier $\Delta = \epsilon-E$ is negative in the case of Fe$^{2+}$
    and positive in the case of Co$^{2+}$. Also, for Fe$^{2+}$ the
    doubly occupied state with energy $E+U$ can be neglected in the
    perturbation theory.}
  \label{fig:energ_diag}
 \end{center}
\end{figure}

In the limit $|\Delta| \gg |t|$ (Fe$^{2+}$), resp. $\Delta,U-\Delta
\gg |t|$ (Co$^{2+}$) the charge transfer between ligands and
metal-ions is suppressed. In order to gain qualitative insight the
fluctuations of the orbital occupation around zero (Fe$^{2+}$) or one
(Co$^{2+}$) can be treated using 2nd order Brillouin-Wigner
perturbation theory (equivalent to a Schrieffer-Wolff transformation~\cite{schrieffer66}). 
We thereby eliminate the charge degrees of
freedom on the metal-ion sites and incorporate their effect in an
effective tunnel coupling between the ligands. An example of a virtual
process giving rise to this coupling is shown in Figure~\ref{fig:energ_diag}. 
In the resulting effective model each metal-ion
site is thus either entirely eliminated (Fe$^{2+}$) or characterized
by a pure spin degree of freedom (Co$^{2+}$). This model contains the
low energy properties of the mobile electrons on the
[$2\times2$]-grid.  As a result we are left with the effective
Hamiltonian (up to a constant):
\begin{eqnarray}
 \label{eq:Fe_eff}
  H_{\mathsf{Fe}}^{\mathsf{eff}} &=& \sum_{\langle j k \rangle}\sum_{\sigma} T\,  a^{\dag}_{j, \sigma} a_{k, \sigma} 
                 + H_{\mathsf{L}} \\
 \label{eq:Co_eff}
  H_{\mathsf{Co}}^{\mathsf{eff}} &=& 
\sum_{\sigma, i} \sum_{j,k=i, i+1 }\big \{(K+J \sigma \tau^{z}_{i}) a^{\dag}_{j, \sigma} a_{k, \sigma} + J \tau^{+\sigma}_{i}   s^{-\sigma}_{j,k} \nonumber\\
&+& (K + (J-F) \sigma \tau^{z}_{i}) n_{j,\sigma} + (J-F)  \tau^{+\sigma}_{i}   s^{-\sigma}_{j,j}\big \} \nonumber \\
 &+& H_{\mathsf{L}} \nonumber \\
&=& \sum_{i} \sum_{j,k=i,i+1}  \big\{(J-F\delta_{j,k}) \bm{\tau}_{i} \, \mathbf{s}_{j,k} + \sum_{\sigma} K
a_{j,\sigma}^{\dag} a_{k, \sigma} \big\} \nonumber \\ &+&  H_{\mathsf{L}} 
\end{eqnarray}
The coupling constants $T$, $K$ and $J$ are 
\begin{eqnarray}
  \label{eq:coupling}
  T &=& \frac{t^{2}}{2 \Delta}\nonumber\\
  K &=& \frac{1}{2} (\frac{t^{2}}{\Delta} - \frac{t^{2}}{U-\Delta}) \nonumber \\
  J &=& \frac{1}{2} (\frac{t^{2}}{\Delta} + \frac{t^{2}}{U-\Delta}).
\end{eqnarray}
For Fe$^{2+}$ the effective Hamiltonian is the extended Hubbard model
 on four ligand sites with effective
hopping matrix element $T$.  In contrast, for Co$^{2+}$ we retain an
eight site model: the effective Hamiltonian couples the spin and
charge on the four ligands to the spin on the four metal-ions. The
first two terms in (\ref{eq:Co_eff}) describe tunneling between
ligands with ($J$) and without spin-flip ($K$). The next two terms
describe fluctuations of the charge and spin on the ligands. The
exchange coupling $J>0$, which is counteracted by the direct
ferromagnetic exchange $F>0$, favors a correlated ground state where
the ligand and metal-ion spins are coupled antiferromagnetically. This
coupling always dominates over the tunneling amplitude $K$: $J > |K|
\geqslant 0$ and $K$ even vanishes for $\Delta =\frac{U}{2}$. The sign
of $K$  depends on whether $\Delta<
\frac{U}{2} (K>0)$ or $\Delta> \frac{U}{2} (K<0)$ and determines
whether the amplitude $K+J$ for the tunneling of electrons with
spin $\sigma$ parallel to the local spin on the metal-ion
$\boldsymbol{S}_{i}$ is enhanced/suppressed relative to the amplitude
$K-J$ for spin antiparallel to $\boldsymbol{S}_{i}$. \\
We point out that the effective Hamiltonian (\ref{eq:Co_eff}) for
Co$^{2+}$ contains no explicit interaction term between the metal-ion
spins: all interactions are mediated by the electrons on the ligands.
In order to describe the electron addition effects on the metal-ion
spin coupling our second-order perturbation theory suffices.  In the
absence of extra electrons on the ligands, only in fourth-order
perturbation theory a weak effective antiferromagnetic Heisenberg
exchange interaction between the spins on the metal-ions appears. This
superexchange is mediated by an empty intermediate ligand orbital. The
four site Heisenberg-model with this effective coupling has been
studied in~\cite{waldmann97} and agrees with intramolecular coupling
found experimentally.  As soon as a ligand orbital contains one
electron, the weak fourth order effect is superseded by the second
order coupling incorporated in the effective Hamiltonian
(\ref{eq:Co_eff}).
\section{Addition energies and spin states\label{sec:Results}}
We now present the results for the effective Hamiltonian
(\ref{eq:Fe_eff}),(\ref{eq:Co_eff}) (perturbation theory) and the full
Hamiltonian (\ref{eq:ham}) .  We first study the addition energy
spectra which reflect mainly the electrostatic effects and then focus
on the spin properties of the ground states and lowest lying excited
states as a function of the number of \textit{added} electrons $n$.
\subsection{Fe$^{2+}$-grid\label{sec:specFe}}
\begin{figure}
 \begin{center}
  \includegraphics[angle=0, scale=0.3]{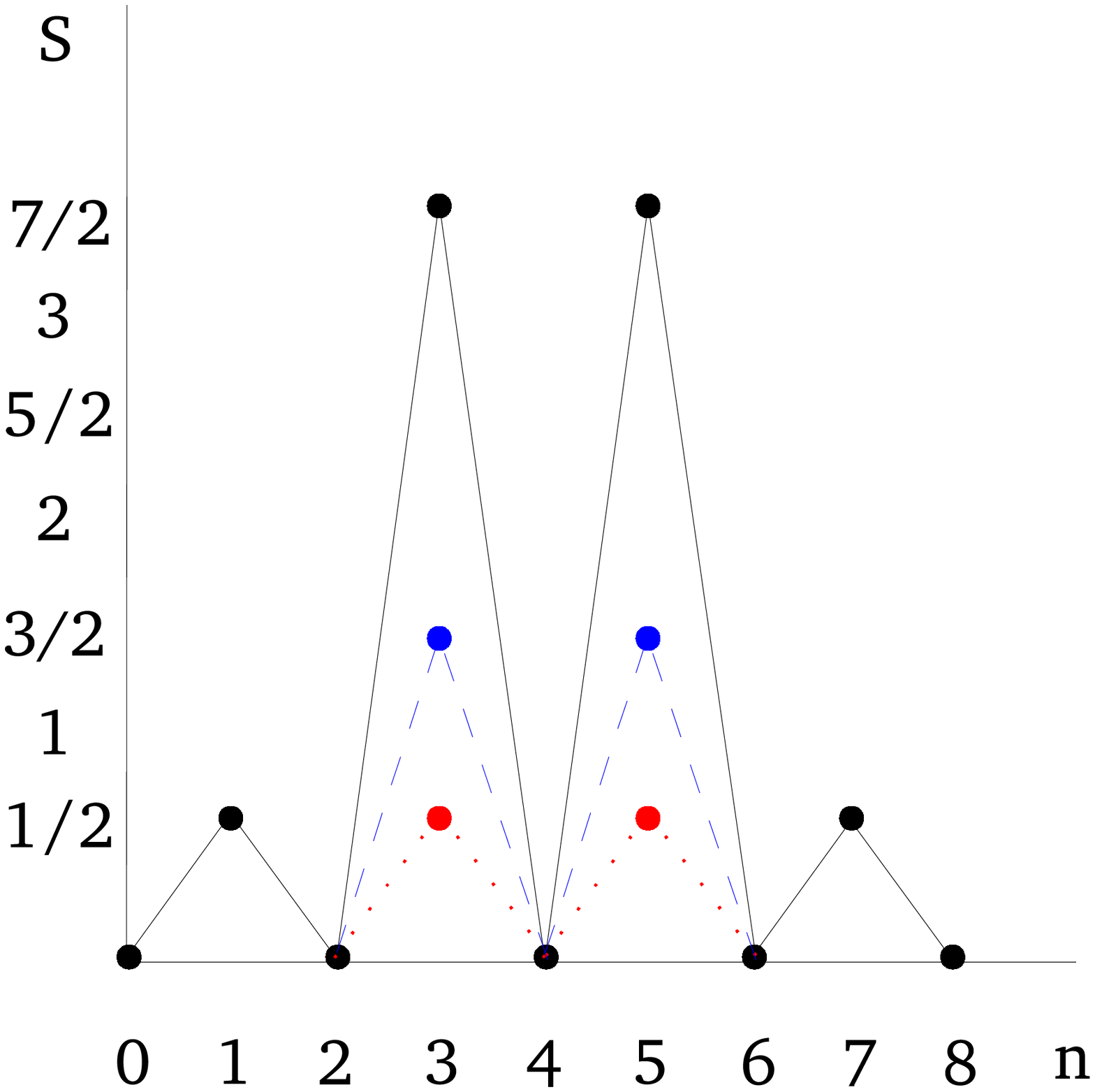}
  \caption{Ground-state spin as function of the number of electrons
    $n$ added to the ligands for $u > u_{\mathsf{th}}$ (dashed blue
    line = Fe$^{2+}$, black line = Co$^{2+}$ without direct exchange) and for $u <
    u_{\mathsf{th}}$ (dotted red line in both cases). }
  \label{fig:gs_spin} 
 \end{center}
\end{figure}

\begin{figure}
 \begin{center}
  \includegraphics[angle=0,  scale=0.3]{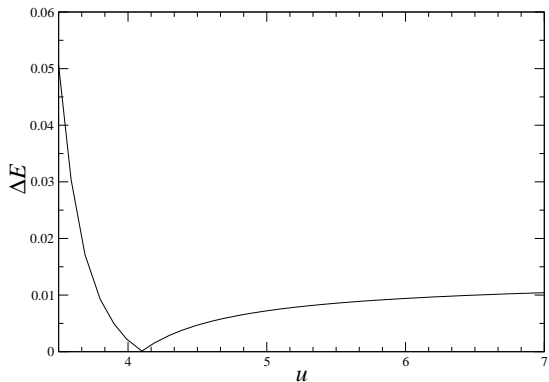}
  \caption{Fe$^{2+}$-[$2\times2$]-grid: Ground-excited state gap as
  function of $u$ for $n=3$, $\Delta=-10,
  v=2.25, w=1$.  For $u > u_{\mathsf{th}} \approx
  4.15$ the ground state has maximal
  spin.}
  \label{fig:splitting}
 \end{center}
\end{figure}

To highlight the effect of the electrostatic interactions, we first
 consider the \textit{noninteracting} limit of the effective 4-orbital
 model (\ref{eq:Fe_eff}) for Fe$^{2+}$ ($u,v,w \ll |T|$).
 Of the four eigenstates the lowest one lies at energy $\epsilon -
 2 |T|$, two states are orbitally degenerate at $\epsilon$ (due to the
 4-fold symmetry axis) and the highest state lies at $\epsilon +
 2 |T|$. Electron addition in this case would give rise to a pair of
 twofold degenerate peaks with a fourfold degenerate peak in
 between. This is in clear qualitative disagreement with the
 experiments ~\cite{ruben03a}.  The addition spectrum can only be
 understood by including the interactions $u > v > w > |T|$ in our
 effective model (\ref{eq:Fe_eff}). The first electron reduces one of
 the four ligands.  The next one goes onto the opposite ligand in
 order to minimize the Coulomb interaction. The third and fourth
 electrons reduce the adjacent ligands. For the next four electrons
 this sequence of processes is repeated, each time doubly occupying a
 ligand orbital.  We thus have two sets of four reduction peaks
 separated by a large gap of order $u$. Each set of four consists of
 two pairs of closely spaced peaks (distance $w$) separated by a
 moderate gap $2v-w < u$. The tunneling between equivalent ligands only weakly affects this picture.
 \begin{figure}
 \begin{center}
  \includegraphics[angle=-90, scale=0.3]{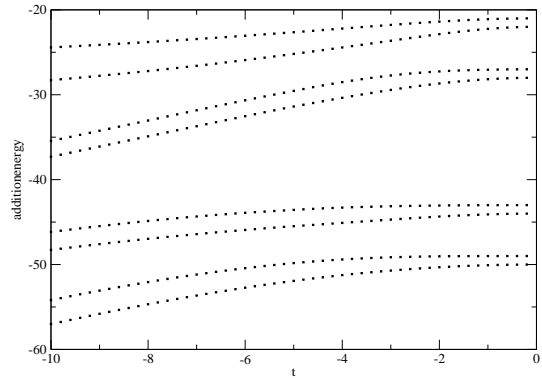}
  \caption{Addition energy of Fe$^{2+}$-[$2\times2$]-grid as function of
 tunneling $t$ (in units of $w$) for $u=15, v=3, w=1, \Delta=-50, U=100$ calculated from
 the full model \ref{eq:ham}. For $|t|<|\Delta|$ we are in the
 perturbative regime where no spin is localized on the Fe$^{2+}$ site
 and the effective model (\ref{eq:Fe_eff}) applies. We reproduce the
 addition energy spectrum measured in~\cite{ruben03a} where the
 spacings between the ground state energies with different $n$ are due
 to electrostatic interactions.}
  \label{fig:add_energy_fe}
 \end{center}
\end{figure}
In Figure \ref{fig:add_energy_fe} we plot the addition energies as a function of the tunneling amplitude
 $t$.\\
Now we discuss the ground state spin as successive electrons are added
to the ligands.  Filling the levels in the \textit{noninteracting}
case of the effective model (\ref{eq:Fe_eff}) ($u,v,w \ll |T|$) according to the Pauli principle the ground state spin is
$S_{\mathsf{tot}}=1/2$ for odd particle number $n$. For even $n=2, 6$
the ground state spin is $S_{\mathsf{tot}}=0$, whereas for
half-filling ($n=4$) $S_{\mathsf{tot}}=0$ and $S_{\mathsf{tot}}=1$ are
degenerate.  In the presence of interactions charge fluctuations are suppressed. For $u \gg |T|$ 
this gives rise at $n=4$ to a Heisenberg antiferromagnet with a
singlet ground state (Figure \ref{fig:gs_spin}).
\\
For \textit{sufficiently large} $u > u_{\mathsf{th}}^{\mathsf{Fe}}$ (Figure
\ref{fig:splitting}) the ground state spin for odd $n=3,5$ is enhanced
from the noninteracting value $S_{\mathsf{tot}}=1/2$ to the maximal
possible value $S_{\mathsf{tot}}=3/2$ (Figure~\ref{fig:gs_spin}).
Now the tunneling between the ligands plays a decisive role.
Because double occupation is suppressed, a single hole/electron (relative to the
half-filled state $n=4$) can maximally gain kinetic energy when the
background of the other electrons is fully spin polarized.  This
ferromagnetic alignment competes with the antiferromagnetic spin
coupling due to superexchange processes. Which process dominates
depends on the strength of the onsite repulsion $u$ relative to the
hopping $|T|$ i.e. $u_{\mathsf{th}}^{\mathsf{Fe}} \propto |T|$.
 The gap between maximal spin ground state and lowest excited state saturates at a
value $\sim 2|T|$ independent of $u$ due to the kinetic origin of the
effect.  This is the underlying mechanism for the Nagaoka theorem
\cite{nagaoka66} which guarantees that the ground state has maximal
spin if $u$ is larger than some positive threshold value. It applies to
the effective model (\ref{eq:Fe_eff}) because it fulfills a certain
connectivity condition for the lattice, namely that a so-called
``exchange loop'' exists which is no longer than four sites
\cite{tasaki89, tasaki98}. This implies that all basis states with
common $S_{\mathsf{tot}}^{z}$ are connected with each other via
nonvanishing matrix elements of (\ref{eq:Fe_eff})~\cite{tasaki98}. In order to
attain an observable effect one should have 
  a moderate  $T$ on the one hand, and on the other hand  the onsite interaction
$u$ must be enhanced  with increasing $T$.
The latter can be achieved by a chemical modification of the ligands that draw charge density in the ligand
LUMO orbitals. Taking typical
parameters~\cite{ruben03a} $ |\Delta| \approx 1 eV, t \approx
10^{-1} eV, \Delta E_{\mathsf{Nag}} \approx 10^{-2}eV, $ we estimate $ u_{\mathsf{th}} \approx
1 eV $ which is reasonable.\\
We have checked that the interactions $v,w$ increase
the critical value $u_{\mathsf{th}}^{\mathsf{Fe}}$ for the Nagaoka state but do not
destroy it~\cite{kollar96}. We have also analyzed the effect of
disorder by making the ligand sites inequivalent through different
MLCT barriers $\Delta$l.
As expected, the Nagaoka state is stable if the change in the MLCT barrier $\Delta$ is smaller than
$|T|$.  Otherwise, due to the localization of electrons the Nagaoka effect, which
is of kinetic nature, is suppressed. We expect disorder effects
 to be relatively weak since the ligands and the metal-ions form
 a highly symmetric grid of equivalent centers.
\subsection{Co$^{2+}$-grid\label{sec:specCo}}
\begin{figure}
 \begin{center}
  \includegraphics[angle=0,scale=0.3]{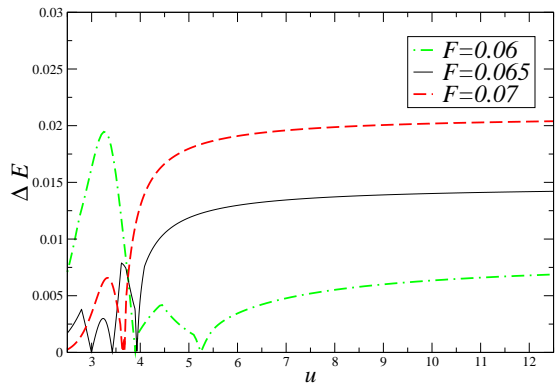}
  \caption{Co$^{2+}$-[$2\times2$]-grid: Ground-excited state gap as
  function of $u$ for $n=3$, $\Delta=10, U=100,
  v=2.25, w=1$ and three different values for the direct exchange $F=0.06,
  0.065, 0.07$.
 The non-monotonic behavior of the gap is due to several level crossings until the 
Nagaoka state is the ground  state (beyond the rightmost zero).
 The direct exchange ``kicks'' the system into
  the Nagaoka state: the threshold value
  $u_{\mathsf{th}}^{\mathsf{Co}}$ is reduced when $F \rightarrow J = 0.06$.
  Even for smaller $F>0$ one still needs  $u>u_{\mathsf{th}}^{\mathsf{Co}}$ to drive the
  system into the maximal spin ground state.
}
  \label{fig:splitting_co}
 \end{center}
\end{figure}

The addition energy spectrum which we obtain for the Co$^{2+}$-grid is
qualitatively the same as for the Fe$^{2+}$-grid since electrostatic
interactions dominate.  In the experiment~\cite{ruben03a} the
reduction peaks corresponding to the first three electrons are similar
to those of Fe$^{2+}$, in agreement with our model.  However, the next
five peaks exhibit roughly a constant spacing $\sim 0.25-0.3 eV$,
corresponding to the charging of one big ``island'' with better
screening.  This would require in our model to artificially change the parameters
to $u \approx v \approx w \approx 0.3 eV$ beyond $n=3$. Obviously effects become
important which are not included in our electronic low temperature
model, e.g.  adding electrons could result in a change in the
molecular geometry which will lead to different electrostatic
interactions. Also at the high experimental temperatures individual
Co$^{2+}$ ions may
be in the high-spin state where the $\pi$-symmetric $t_{2g}$-orbitals are singly
occupied. These can couple more strongly to the ligand orbitals and
increase metal-ligand charge transfer.
Here we are interested in the
low temperature regime however and assume low-spin ($S=1/2$) Co$^{2+}$ ions.\\
The spin properties of the eigenstates of the effective model
(\ref{eq:Co_eff}) without direct exchange interaction, i.e. $F=0$, are
qualitatively similar to Fe$^{2+}$ (Figure~\ref{fig:gs_spin})
and will be considered first.  We
have a singlet ground state at half-filling ($n=4$), and a Nagaoka
maximal-spin ground state near half-filling ($n=3,5$) for sufficiently
large charging ($U > u>  u_{\mathsf{th}}^{\mathsf{Co}}$). At half-filling $n=4$ the
antiferromagnetic N{\'e}el-state has the largest weight in the ground
state.  The electron spins on the metal-ions couple ferromagnetically
due to the presence of electrons on the ligands and \textit{and vice
versa}.  This is to be contrasted to the situation at $n=0$ where the
metal-ion spins couple antiferromagnetically and at $n=4$ for the
Fe$^{2+}$-[$2\times2$]-grid, where the electron spins on adjacent
ligands couple antiferromagnetically.  The total spins of the
metal-ion and ligand sublattice couple antiferromagnetically to a
singlet ground state.  The appearance of the Nagaoka state at $n=3,5$
has a different origin than in the Fe$^{2+}$-[$2\times2$]-grid since
we have exchange loops longer than four sites. We do, however, have a
bipartite lattice and hopping occurs only between the ligand and
metal-ion sublattices. This is also a sufficient condition for the
Nagaoka theorem to apply~\cite{kollar96,tasaki98}.\\
The four spins on the bridging Co$^{2+}$ metal-ions cause  two
quantitative differences from the case of Fe$^{2+}$.  Firstly, the
maximal spin value attained in the Nagaoka state is simply larger,
$S_{\mathsf{tot}}=7/2$.  Secondly, the threshold value of
$u_{\mathsf{th}}^{\mathsf{Co}}$ for the appearance of this state is dramatically
increased, $u_{\mathsf{th}}^{\mathsf{Co}} \approx 10^{4}$ in both the full and
effective models. This is due to the antiferromagnetic exchange
coupling $J$ between the metal-ion and ligand sublattices.
This suppression of ferromagnetism can however be understood by
considering two extreme limits. In the limit where two sublattices are
equivalent ($\Delta=0, u=U$) Nagaoka's mechanism is ineffective
because the exchange paths are to long ($> 4$ sites,~\cite{tasaki98}). This implies that with
decreasing $\Delta$ a higher $u_{\mathsf{th}}^{\mathsf{Co}}$ is required to
stabilize the Nagaoka state.  In the opposite limit where the
sublattices are well separated in energy ($\Delta, U-\Delta \gg |t|$),
the effective model (\ref{eq:Co_eff}) applies and  $u_{\mathsf{th}}^{\mathsf{Co}}$ is decreased. 
The maximal spin state can still be achieved when $U$ is sufficiently
increased for \textit{fixed} $\Delta$, such that $t \ll \Delta \ll
U/2$ we have $K \approx - J$: tunneling of electrons with spins
antiparallel to the metal spins is suppressed, favoring a polarized
ground state.
However, the splitting $\Delta E$ is decreased due to the small
effective coupling constants.\\
The neglect of any (even small) direct exchange represents an
unbalanced treatment of the magnetic coupling, since the Nagaoka
mechanism and the direct exchange are known to cooperate to stabilize
maximal spin states~\cite{kollar96}. Even when it is weak, $F \approx
J$, the direct exchange leads to a dramatic reduction of
the threshold Coulomb energy for achieving the Nagaoka state to values
more close to those found above for Fe$^{2+}$: $u_{\mathsf{th}}^{\mathsf{Co}}\sim 5$ 
(Figure~\ref{fig:splitting_co}) and the gap between ground and excited
state increases again.
 Furthermore it is possible that an enhanced direct 
exchange stabilizes a maximal spin state even at half-filling $n=4$. 
Still the spin can be switched from $0$ to $7/2$ by going from $n=2$
to $n=3$ (or from $n=5$ to $n=6$).
The excitation gap $\Delta E$ as function of $u$ (Fig.~\ref{fig:splitting_co}) 
saturates at values of the order of $0.2 J$.  In this limit
the direct exchange ``kicks''~\cite{kollar96} the system into the
Nagaoka state.  For $F \gg J$ the ferromagnetic coupling exceeds the
exchange coupling and the resulting ground state is always
ferromagnetic, independent of $u$ indicating the Nagaoka mechanism is
not relevant anymore. 
\section{Conclusion\label{sec:disc}}
We have analyzed a strongly-correlated electron model for a
[$2\times2$]-grid complex with four transition-metal centers (either
Fe$^{2+}$ or Co$^{2+}$) to illustrate the interplay between
electron addition and intramolecular spin coupling.
 Our model contains both localized
magnetic moments and delocalized electrons, in contrast to the customary
description of molecular magnets.  We have based our model on the
addition energy spectra of the experiments in
Ref.\cite{ruben03a}, the crucial input being that the extra electrons
occupy ligand orbitals. We found large changes in the total spin, $\Delta S_{\mathsf{tot}} > 1/2$, of the
 molecule upon variation of the total electron number due to the
 Nagaoka mechanism. The large charging energies on the ligands required for the high spin
 states can be tuned chemically by adding electron-donating groups to
 the ligands. Localized spins on the mediating metal ions (Co$^{2+}$) counteract
 the Nagaoka effect, but the direct exchange coupling with the
 neighboring ligands can compensate for this.
 The total spin in the Nagaoka state for the Co$^{2+}$ complex is
 therefore more than twice as large as for the Fe$^{2+}$ complex.
 Low temperature electron tunneling experiments can access the change
 of the molecular spin as a function of added charge.
 Spin-blockade effects~\cite{weinmann95} will dominate the single-electron tunneling around transitions between
 charge states with maximal spin~\cite{romeike05a}.
 Also, the $S_{\mathsf{tot}}=1/2$ Kondo effect usually expected for odd $n=3,5$ will
 also be suppressed.  In any case the Nagaoka mechanism lowers maximal spin states in
 energy. Even as low lying total-spin excitations Nagaoka states have
 clear transport fingerprints due to spin-selection rules~\cite{romeike05a}.
 Other molecular complexes can also show the above behavior.
 It is essential that the connectivity of the electron-accepting centers
 is appropriate~\cite{kollar96,tasaki98} for the Nagaoka mechanism
 to be effective.
\begin{acknowledgments}
J. Kortus is acknowledged for stimulating discussions. We thank
J.-M. Lehn for providing us experimental data and for discussions. M. R. Wegewijs
acknowledges the financial support provided through the European 
Community's Research Training Networks Program under contract
HPRN-CT-2002-00302, Spintronics.
\end{acknowledgments}

\end{document}